\documentclass[usenatbib]{mn2e}
\usepackage{times}
\usepackage{epsfig}
\usepackage{colordvi}

\usepackage{graphicx}
\usepackage{amssymb}

\def\araa{ARA\&A}%
\def\apj{ApJ}%
\def\apjl{ApJ}%
\def\aap{A\&A}%
\def\mnras{MNRAS}%
\def\pasj{PASJ}%

\def\Rs{R_{\rm S}}
\def\Rns{R}
\def\Mns{M}
\def\msun{{\rm M}_{\odot}}
\def\Tc{T_{\rm c}}
\def\Tcr{T_{\rm Edd}}
\def\Teff{T_{\rm eff}}
\def\TS{T_{\rm S}}
\def\Ledd{L_{\rm Edd}}
\def\gwr{g_{\rm cor}}
\def\grad{g_{\rm rad}}
\def\geff{g_{\rm eff}}
\def\Hbl{H_{\rm BL}}
\def\hbl{h_{\rm BL}}
\def\Lbl{L_{\rm BL}}
\def\fc{f_{\rm c}}
\def\sigmasb{\sigma_{\rm SB}}
\def\sigmae{\sigma_{\rm e}}
\def\taue{\tau_{\rm e}}
\def\me{m_{\rm e}}

\def\hs{h_{\rm S}}
\def\rhos{\rho_{\rm S}}
\def\Sigmas{\Sigma_{\rm S}}

\def\vphi{v_{\varphi}}
\def\vtheta{v_{\theta}}
\def\fphi{f_{\varphi}}
\def\ftheta{f_{\theta}}
\def\fr{f_r} 
\def\tauphi{\tau_{\varphi}}
\def\tautheta{\tau_{\theta}}
\def\alphab{\alpha_{\rm b}}

\def \Omegak{\Omega_{\rm K}}
\def \Omegans{\Omega_{\rm *}}

\def\vecv{\bmath{v}}
\def\vecq{\bmath{q}}
\def\vecf{\bmath{f}}

\def\d{{\rm d}}

\newcommand{\be}{\begin{equation}}
\newcommand{\ee}{\end{equation}}

\title[Spectra of neutron star spreading layers]
{Spectra of the spreading layers on the neutron star surface and 
 constraints on the neutron star equation of state}

 \author[Valery Suleimanov \& Juri Poutanen]
    {Valery~Suleimanov$^{1,2,3}$       and    Juri~Poutanen$^{1}$ \\
$^1$Astronomy Division, PO Box 3000, FIN-90014 University of Oulu,
            Finland \\
$^2$Institut f\"ur Astronomie und Astrophysik, Abteilung   Astronomie,
          Sand 1, D-72076 T\"ubingen, Germany\\
$^3$Kazan State University, Kremlevskaya str.18,     420008 Kazan, Russia}

\begin{document}
\date{Accepted, Received}
\pagerange{\pageref{firstpage}--\pageref{lastpage}} \pubyear{2005}
\maketitle

\label{firstpage}

\begin{abstract}
Spectra of the spreading layers on the neutron star surface are calculated
on the basis of the Inogamov-Sunyaev model taking into account general
relativity correction to the surface gravity and considering various 
chemical composition  of the accreting matter. 
Local (at a given latitude) spectra are  similar
to the X-ray burst spectra and are described by a diluted black body. Total
 spreading layer spectra are integrated accounting for the light bending,
gravitational redshift, and the relativistic Doppler effect and aberration.
They depend slightly  on the inclination angle  and on the luminosity.  
These spectra also can be fitted by a diluted black body with the color
temperature depending mainly on a neutron star compactness.
Owing to the fact that the flux from the spreading layer is close to the critical 
Eddington, we can put constraints on a neutron star radius without the need 
to know precisely the emitting region area or the distance to the source.
The  boundary layer spectra observed in the luminous low-mass X-ray binaries,
and described by a black body of color temperature $\Tc=2.4 \pm 0.1$ keV, 
restrict  the neutron star radii   to $R=14.8 \pm 1.5$ km (for a 1.4-$\msun$ 
star and solar composition of the accreting matter), which
corresponds to the hard equation of state.
\end{abstract}


\section{Introduction}

Matter accreting on to a weakly magnetized neutron star (NS) in low mass X-ray
binaries (LMXRBs) can form an accretion disc which extend down to the
NS surface.  A boundary layer (BL) 
is formed between the accretion disc and the NS surface, 
where a rapidly rotating matter of the disc is decelerated down to the
NS angular velocity.  The amount of the energy, which is
generated during this process, is comparable with the energy generated in
the accretion disc \citep{SS86,SS98}.

There is no generally accepted theory of the BL. Two different approaches to
the BL description are considered. First of them, which we will call
a `classical model', considers the BL between a central star (a white
dwarf or a NS) as a part of the accretion disc
\citep{P77,PS79,T81,SS88,BK94,PN95,PS01}.
In this model the component of velocity normal to the accretion disc plane
 is zero. The half-thickness of the BL is determined by the same relation,
as for the accretion disc:
\be \label{eq:Hbl}
     \Hbl\sim \frac{c_{\rm s}}{v_{\rm K}} R,
\ee
where $c_{\rm s}$ is a sound speed in the BL and $v_{\rm K}$ is the
Keplerian velocity close to the NS surface of radius $R$.
The radial extension
of the BL is determined by the relation \citep{P77}
\be \label{eq:hbl}
     \hbl\sim \frac{c^2_{\rm s}}{v^2_{\rm K}} R
     \sim \Hbl\frac{\Hbl}{R}.
\ee
In this classical model the accreting matter in the BL
is decelerated in the accretion disc plane, along radial coordinate only,
due to the viscosity operating within the differentially rotating BL,
similarly to the accretion disc. From the observational point
of view, the classical BL is a bright equatorial belt close to the
NS surface.  The effective temperature of the BL is higher than
the maximum accretion disc effective temperature, because the BL is
smaller than the accretion disc, while their luminosities  are comparable.

Another approach was suggested by  \citet[][ hereafter IS99]{IS99}.
The BL is considered as a spreading layer (SL)
on the NS surface. The accreting matter diffusing along
the radial direction in the accretion disc  and reaching the neutron
star surface gains the velocity component normal to the accretion disc
plane due to the ram pressure from  the accretion disc. Then the
matter spirals along the NS surface  towards the
poles.  Rotating at the NS surface, the matter is decelerated due
to a turbulent friction between the rapidly rotating matter and a slowly
rotating NS surface. The kinetic energy of the accreting gas
is mostly deposited in
two bright belts at some latitude above and below the NS equator.
The width and the latitude of the belts depend on the mass accretion rate. The larger
the accretion rate, the wider the belts are  and the closer they are to the NS
poles.  At the accretion rate close to the Eddington limit ($L_{\rm BL}
\sim \Ledd$) the bright belts  expand all over the NS surface.

The observational difference between two BL models is not significant.
At low accretion rates ($\Lbl\sim 0.01 \Ledd$)
the latitude of bright belts of the SL is small ($\sim$ few
degrees) and the vertical extension of the SL is comparable
to the classical BL thickness. At high accretion
rates ($\Lbl\sim  \Ledd$) the classical BL thickness is
comparable to the NS radius   \citep[see][]{PS01}.
Therefore, the effective temperatures of these BL models are of the
same order. 

In the approach by IS99, the NS radius is assumed to be larger that $3\Rs$ 
(where $\Rs=2GM/c^2$ is the Schwarzschild radius of a NS of mass $M$), 
but the accretion disc structure is not significantly changed up to the NS
surface, and the radial velocity is always subsonic.
If the radial velocity of accreting gas is supersonic at the surface 
(see e.g.  \citealt*{PN92}, for a possibility of the supersonic radial velocity    
in classical BL, and \citealt*{KW91}, for the ``gap accretion'' when 
the NS is within the innermost stable circular orbit), 
some  fraction of the kinetic energy (associated with 
a small radial velocity component) should be dissipated in an oblique shock, but 
most of it  still remains stored in the kinetic energy of the gas rotating  at the surface 
to be dissipated later  in the SL. 
The gap accretion  model of \citet{KW91} is rather similar to the SL, but they 
did not consider  the fate of the accreting material and its spread over the surface. 
At  very low accretion rate, both models 
should produce hard Comptonization spectra extending to $\sim100$ keV. 

The aim of this work is the calculation of the radiation spectra of the
SLs  and their comparison to the observed
X-ray spectra of the BLs in the luminous LMXRBs.

\section{Spreading layer model}

The theory of the SL was developed by IS99
 under the assumption of Newtonian gravity. They
considered accretion of the pure hydrogen plasma. Here we
repeat the IS99 theory for plasmas of arbitrary chemical composition
taking into account  general relativity (GR) corrections using the
pseudo-Newtonian potential. These corrections may be important, because
the maximum effective temperature of the SL, which should be
smaller than the local Eddington effective temperature $\Tcr$,
depends on the opacity and the gravity. The critical
temperature is determined by the balance between the surface gravity and
the radiative acceleration:
\be \label{eq:tc}
        \frac{G \Mns}{\Rns^2 \sqrt{1-\Rs/\Rns}} = \frac{\sigmasb \Tcr^4}{c}
        \sigmae,
\ee
where $\sigmae = 0.2 (1+X)$ cm$^2$~g$^{-1}$ is the electron scattering opacity, $X$ is the hydrogen mass fraction, and   $\sigmasb$ is the Stefan-Boltzmann constant. 
It is clear, that solar (or larger) helium
abundance together with the GR correction will lead to higher $\Tcr$, 
and, therefore, to a higher maximum effective temperature of the SL.
This could have important consequences for the determination of the neutron
star parameters from observations.

We use the pseudo-Newtonian potential in the form:
\be \label{eq:fi}
    \Psi (r) = -c^2 \left(1 - \sqrt{1-\Rs/r}\right).
\ee
    This potential gives the correct GR  surface gravity
    \be \label{eq:g0}
    g_{\rm 0}(\Rns)= \frac{G \Mns}{\Rns^2\sqrt{1-\Rs/\Rns}},
    \ee
but gives  the Keplerian velocity at
the NS surface
    \be \label{eq:vk}
    v^2_{\rm K}(\Rns) = \frac{G \Mns}{\Rns\sqrt{1-\Rs/\Rns}},
    \ee
    which is smaller  than the correct GR value.

\subsection{Main equations}

\begin{figure}
\centerline{ \epsfig{file=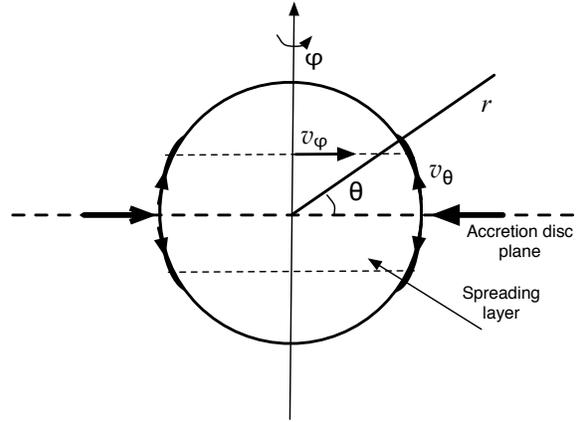,width=8cm}}
\caption{\label{fig1} Geometry of the problem.}
\end{figure}

Below we rewrite the IS99 equations for the SL for the pseudo-Newtonian
potential (\ref{eq:fi}) and considering  arbitrary abundances.
We consider the dynamics of the SL on the NS
surface (see Fig.~\ref{fig1}). The full hydrodynamic equations, which
describe this process are as follows \citep[see for example][]{M78}.  The
continuity equation is
\be \label{eq:cont}
 \frac{\partial \rho}{\partial t} + {\bf \nabla \cdot} (\rho \vecv) =0,
\ee
where $\rho$ is the plasma density, $\vecv$ is the vector of the
gas velocity in the SL with components $\vphi$, $\vtheta$
and
$v_{r}$, which are velocities of the SL along longitude, latitude and
radius correspondingly.
Conservation of momentum for each gas element is described by the
vector Euler equation
\be \label{eq:euler}
\rho \frac{\partial \vecv}{\partial t} + 
\rho  \vecv \cdot \nabla \vecv = - \nabla P +  \vecf,
\ee
where $P=P_{\rm rad}+P_{\rm g}$ is the total pressure which is a sum of the radiation
and gas pressures, and $\vecf$ is a force density.  The energy equation
for the gas in the SL is
\begin{eqnarray} \label{eq:energy}
\frac{\partial}{\partial t} \left( \frac{1}{2} \rho v^2
 +\varepsilon\right) +  \nabla \cdot 
\left[ \left(\frac{1}{2} \rho v^2 + \varepsilon + P\right){\vecv}\right] =
\\ \nonumber  \vecf \cdot \vecv - \nabla \cdot \vecq +Q^+.
\end{eqnarray}
Here $\varepsilon = \varepsilon_{\rm rad}+ \varepsilon_{\rm g}$ is the total
density of internal energy, where $\varepsilon_{\rm rad}=aT^4$ is the
radiation energy density and $\varepsilon_{\rm g} = (3/2) P_{\rm g}$ is the density
of the internal gas energy. The first term on the right hand side 
is the power produces by the force density, the second is
the energy, which is lost by radiation ($\vecq$ is a vector of the
radiation flux), and the third is the heat, which is generated within a
unit volume of the SL.

Following IS99 we consider the steady state SL model in the
spherical coordinate system $(r,\theta,\varphi)$, where $\theta$
is the latitude and $\varphi$ is the azimuthal angle (see Fig.~\ref{fig1}).
We also assume that the SL has a
small thickness (in comparison with the NS radius $\Rns$, therefore the
radial coordinate $r=\Rns$), the radial velocity component
is zero $v_r=0$, and it is axially symmetric
(therefore, all of the derivatives $\partial / \partial \varphi$ equal to
zero).  In this case
equations (\ref{eq:cont})--(\ref{eq:energy}) take the following form.
The continuity equation is
\be  \label{eq:cont1}
\frac {1}{R \cos \theta} \frac{\partial}{\partial \theta} (\cos \theta\  \rho
\vtheta) = 0 ,
\ee
the
three components of the Euler equation are
\be  \label{eq:e1}
     -\rho\left(\frac{\vtheta^2 +\vphi^2}{R}\right) =
-\frac{\partial P}{\partial r} +\fr,
\ee
\be  \label{eq:e2}
    \rho \frac{\vtheta}{R} \frac{\partial \vtheta}{\partial
\theta} + \rho \frac{\vphi^2}{R} \tan \theta =
-\frac{1}{R}\frac{\partial P}{\partial \theta} +\ftheta,
\ee
\be  \label{eq:e3}
    \rho \frac{\vtheta}{R} \frac{\partial \vphi}{\partial
\theta} - \rho \frac{\vphi \vtheta}{R} \tan \theta =
\fphi,
\ee
and the energy equation is
\begin{eqnarray} \label{eq:energy1}
    \frac{1}{R \cos \theta} \frac{\partial}{\partial \theta} 
    \left[ \cos\theta \  \vtheta
\left(\frac{1}{2} \rho v_0^2 + \varepsilon +P\right)\right] =
\\ \nonumber
\ftheta\vtheta + \fphi\vphi - \frac{\partial q}{\partial r} + Q^+,
\end{eqnarray}
where
\be  \label{v0}
 v_0^2= \vphi^2 + \vtheta^2.
\ee
Here the radiation flux has only one (radial) non-zero
component and   its divergence is computed in the plane-parallel approximation
which is a consequence of our assumption of small height of the SL.
A small azimuthal component of the radiation flux  arises due to the aberration,
which we neglect here.

It is clear that equation~(\ref{eq:e1}) can be solved independently on equations
(\ref{eq:e2})--(\ref{eq:e3}) and we can consider some averaging over the
layer's height. Therefore, we  arrive at a one-dimensional problem.
In this case  instead of equations (\ref{eq:cont1})--(\ref{eq:energy1})
we obtain
\be  \label{eq:cont2}
\frac {1}{R \cos \theta} \frac{\partial}{\partial \theta} \left( 
\cos \theta \int \rho \vtheta \d r\right)  = 0 ,
\ee

\be  \label{eq:e22}
  \int  \rho \vtheta \frac{\partial \vtheta}{\partial \theta} \d  r + \tan \theta \int \rho \vphi^2 \d  r 
  = -\frac{\partial}{\partial \theta} \int P \d  r + R \int \ftheta \d  r ,
\ee

\be \label{eq:e32}
\int \rho \vtheta
\frac{\partial \vphi}{\partial \theta} \d  r - \tan \theta
\int \rho \vphi \vtheta \d  r  = R \int  \fphi \d  r ,
\ee

\begin{eqnarray} \label{eq:energy2}
    \frac{1}{R \cos \theta} \frac{\partial}{\partial \theta}
     \left[ \cos\theta \int \vtheta
\left(\frac{1}{2} \rho v_0^2 + \varepsilon +P\right) \d  r \right] =
\\ \nonumber
\int \ftheta\vtheta \d  r + \int \fphi\vphi \d  r -  q + \int Q^+ \d  r. 
\end{eqnarray}
Here the integration over radius is from $R$ to $R+\hs$,
where $\hs$ is the local SL thickness. We define the corresponding
force densities   in the next section.

\subsection{Vertical averaging}
\label{sec:avehei}

The one-dimensional equations for the SL structure are derived using
the averaging along the height at a given latitude. It means that we have
to calculate all   the integrals in   equations
(\ref{eq:cont2})--(\ref{eq:energy2}) for some model of the SL vertical structure.
The simplest way is just to consider   the variables averaged over the height.

IS99 used a
more complicated model of averaging. They constructed  a simple model
of the SL   using assumptions  that velocities $\vtheta$, $\vphi$ and the radiation flux 
do not depend on the height $q(r)=\mbox{const}=\sigmasb \Teff^4$.
The latter suggestion means that all of the thermal
energy in the SL is generated at the bottom. This model is described by the
hydrostatic equilibrium equation (\ref{eq:e1}) taken in the form
\be  \label{eq:he}
    \frac{\d P}{\d m} =  \geff \equiv  g_0 - \frac{\vphi^2+ \vtheta^2}{R},
\ee
and the radiation transfer
equation in the diffusion approximation
\be \label{eq:re}
    \frac{\d\varepsilon_{\rm rad}}{\d m} = \frac{3q}{c} \sigmae.
\ee
Here and below we use a new independent variable: a column mass
$m$ and a new geometrical coordinate $z$, which are defined as
\be
\d m = \rho \d z     = -\rho \d r.    \nonumber
\ee
Coordinate $z$ has an opposite direction relative to $r$ and $z$=0 at
$r=R+\hs$. We also defined the $r$ component of the force density
\be
\fr = -g_0 \rho.   \nonumber
\ee
Equations (\ref{eq:he})--(\ref{eq:re}) have to be supplemented by the equation
of state
\be \label{eq:se}
P = \frac{\rho kT}{\mu m_{\rm p}} +
\frac{\varepsilon_{\rm rad}}{3},
\ee
where $\mu=4/(3+5X)$ is the mean molecular
weight and $m_{\rm p}$ is the proton mass.

Equations (\ref{eq:he})--(\ref{eq:se}) can be solved with the simple
boundary conditions $P(m=0)=0$, $T(m=0)=0$:
\be
     P=\geff m,
\ee
\be
     \varepsilon_{\rm rad}= \frac{3q}{c} m \sigmae,
\ee
\be
     \rho = \mu m_{\rm p} \frac{\gwr}{k} \left( \frac{a c}{3q
\sigmae}m^3 \right)^{1/4},
\ee
\be
     T=\left( \frac{3q}{ac} m \sigmae \right)^{1/4}= \Teff
        \left( \frac{3}{4} \taue\right)^{1/4},
\ee
where $\taue= m \sigmae$ is the electron scattering optical
depth of the layer, and   
\be
      \gwr \equiv \geff - \grad = \geff - \frac{q}{c}
      \sigmae.
\ee
The column density $m$ is related to the geometrical depth~$z$
\be
      m = \frac{(\mu m_{\rm p} \gwr)^4}{4^4 \sigmae k^4}
           \frac{ac}{3q} z^4,
\ee
which gives the following dependence of temperature on height
\be
      T=\frac{\mu m_{\rm p} \gwr}{4k} z.
\ee

Following IS99, we consider  the values of temperature and density at the
bottom of the SL  $\TS$  and $\rho_{\rm S}$ as  parameters. 
In this case the local SL thickness $\hs$ is:
\be  \label{eq:height}
      \hs =
      \frac{4k\TS}{\mu m_{\rm p} \gwr}.
\ee
We can  also calculate all of the integrals in
equations~(\ref{eq:cont2})--(\ref{eq:energy2}): the total surface density
\be
     \int_{0}^{\hs} \rho \d z = m(z=\hs) = \Sigmas,
\ee
the pressure  surface density
\be
\int^{\hs}_0 P  \d z  = \frac{1}{5}
\geff \hs \Sigmas= \frac{4}{5} \frac{\geff}{\mu
m_{\rm p}\gwr} \Sigmas k \TS,
\ee
the surface density of the internal energy
\begin{eqnarray}
   E_{\rm int} = \int^{\hs}_0 \left(\varepsilon_{\rm rad} +
\frac{3}{2}
P_{\rm g}\right)  \d z  =
\frac{3}{2} \left(\geff+\grad\right)
\frac{\Sigmas \hs}{5},
\end{eqnarray}
the local flux
\be
   q=\sigmasb \Teff^4  = \frac{ac}{3\sigmae}
\frac{\TS^4}{\Sigmas},
\ee
and the enthalpy flux
\be
H= E_{\rm int} + \int^{\hs}_0 P  \d z =
 \left( \frac{5}{2} \geff + \frac{3}{2} \grad\right)
\frac{\Sigmas \hs}{5} .
\ee
If we   take $X=1$, all these relations will be the same, as
derived by IS99 with one exception:  there is no
potential energy of the SL in our energy equation. 
Thus our expression for $H$ contains a factor
$5/2$ instead of $7/2$ as in IS99. Below we will show that this
produces only a small quantitative differences between our
and IS99 models.

In the IS99 model there are two forces, which give contribution to
the force density in  equations (\ref{eq:cont2})--(\ref{eq:energy2}). These
are the gravity force, which has only the radial component (see above) and the
force arising due to the friction between the NS surface and
the SL.  This force is directed along the NS surface and is
expressed in the IS99 model through stress  $\tau$ and its azimuthal and
meridional components
$\tauphi$ and  $\tautheta$.
IS99 have parameterized it in the form:
\begin{eqnarray}
\tauphi&=& -\int_0^{\hs} \fphi\d z =  \alphab
\rhos\vphi v_0,\\ \nonumber
\tautheta &=& -\int_0^{\hs} \ftheta\d z =  \alphab
 \rhos\vtheta v_0,
\end{eqnarray}
where $\alphab=v_*^2/v_0^2$  is the parameter of the stress  parametrization,
$v_*$ is the velocity of turbulent pulsations. 
We should note that $\alphab$ is not the same $\alpha$ that is
used in the accretion disc theory. In accretion discs, $\alpha$
(in the first approximation) is the square of the ratio of the
turbulent velocity to the sound speed  $\alpha = v_*^2/c_{\rm s}^2$ and 
can be quite high, up to 0.1--1.  
In the SL, the plasma velocity $v_0$ is close to the Keplerian velocity 
at the NS  surface and is orders of magnitude larger than the  sound speed.
The velocity of turbulent pulsation is also limited by  the radiation
viscosity at the SL bottom. IS99 carefully investigated this matter  
and estimated $\alphab \sim 10^{-3}$. 
We used this value in most of the paper.

IS99 have ignored the mechanical work between the SL and the NS (which
accelerates the stellar rotation). In our work we use the same approximation.
A fraction  of the kinetic energy of the accreting gas that goes to 
increase the rotational energy of the NS is approximately $2 \Omegans/\Omegak$,
where $\Omegans$ is the NS angular velocity and $\Omegak$ is the Keplerian 
angular velocity at the NS surface. 
As we consider a non-rotating NS and the characteristic time to increase 
its angular velocity is orders of magnitude larger than the
characteristic time of the SL  $t= \Rns/\vtheta=10\ \mbox{km}/ 10^3\ \mbox{km s}^{-1} = 0.01$ s,
ignoring the mechanical work on the NS is a reasonable approximation. 
In this approximation, therefore, all the work due to friction between the SL and the 
NS transforms to heat:
\be  \label{eq:heat}
    \int_0^{\hs} Q ^+ \d z = - (\tauphi\vphi +
\tautheta \vtheta) = -\tau v_0.
\ee

\subsection{One-dimensional model of the spreading layer}
\label{sec:structure}

 Using relations (\ref{eq:height})--(\ref{eq:heat}) we can rewrite equations,
which describe the one-dimensional SL structure.  The
continuity equation can be rewritten via the accretion rate as 
\be \label{eq:cn}
     \dot{M} = 4 \pi R \ \cos\theta\ \vtheta \Sigmas,
\ee
Therefore, the product $\cos\theta \ \vtheta \Sigmas =const$. The Euler
equations are
\begin{eqnarray} \label{eq:fine1}
    \cos\theta\ \vtheta \Sigmas \vtheta' + \frac{4}{5} \cos\theta\
    \left( \frac{\geff}{\gwr}\frac{k\TS}{\mu m_{\rm p}}
\Sigmas\right)' = \\ \nonumber
-R\cos\theta\  \tautheta - \sin\theta\ \Sigmas \vphi^2 ,
\end{eqnarray}

\be  \label{eq:fine2}
     \Sigmas\vtheta(\cos\theta\ \vphi)' =   -R \cos\theta \tauphi.
\ee
    Here the prime means the derivative over $\theta$.
The second term in the left hand side of   equation~(\ref{eq:fine1}) is the
lateral force gradient, and the terms in the right hand side are the
components of the stress force and the centrifugal force. We
see from equation~(\ref{eq:fine2}) that the momentum along $\varphi$
coordinate is changed due to the friction with the NS surface
only.  The system of equations is closed by the energy equation
\be \label{eq:en}
    \Sigmas\vtheta \left( \frac{v_0^2}{2} + \frac{2}{5}
\frac{k\TS}{\mu m_{\rm p}}\frac{5\geff+3\grad}{\gwr}
\right)' = -R q .
\ee

\begin{figure}
\centerline{ \epsfig{file=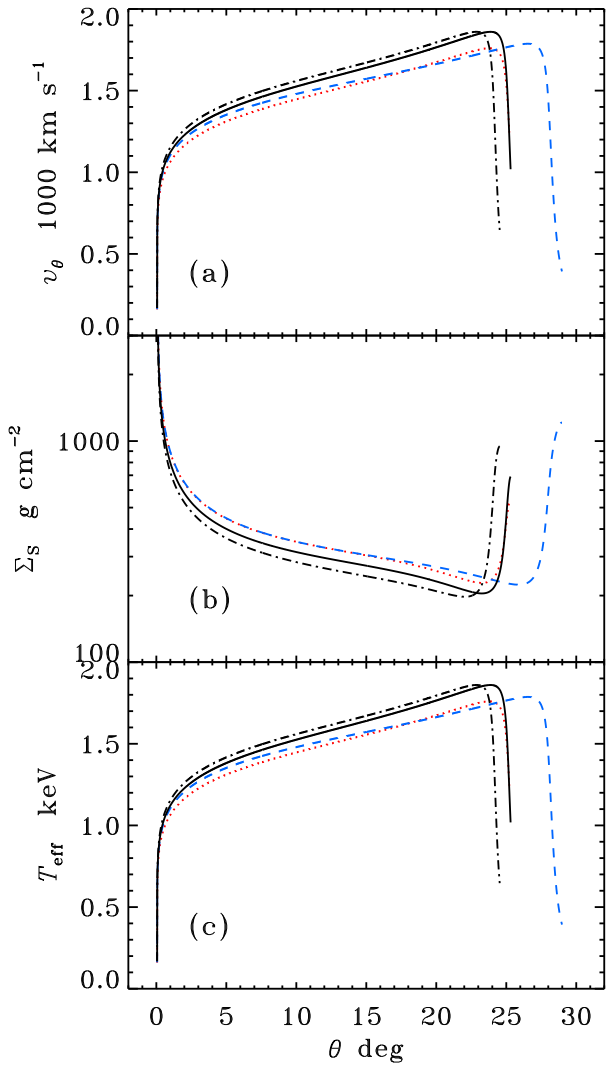,width=8cm}}
\caption{\label{fig2}
The distributions of  (a) $\vtheta$, (b) the surface density
$\Sigmas$  and (c) the effective temperature $\Teff$  along
latitude $\theta$ for four models.
Solid  and dashed curves are for our model
corrected for the GR  using pseudo-Newtonian potential and $X=0.7$ and $X=1$,
respectively. The
dotted curves is our model  for the Newtonian gravity and $X=0.7$, while the
dot-dashed curves are for the IS99 model corrected to GR  and $X=0.7$.
In   all cases $\Mns = 1.4
\msun$ and $\Rns$ = 12 km. The accretion rate of all models
is the same, corresponding to $0.2 \Ledd$ of the first model.}
\end{figure}

The system of equations~(\ref{eq:cn})--(\ref{eq:en}) can be transformed to the three
dimensionless equations  for $\vphi(\theta)$, 
$\vtheta(\theta)$ and $\TS (\theta)$ as was done by IS99.
These equations are solved with the boundary conditions at
the transition zone between the accretion disc and the SL: the initial
latitude, where the SL starts, is close to the
NS equator $\theta_{\rm 0} \sim $ 10$^{-2}$;
 the initial relative
deviation $\delta$ of $\vphi$ from the Keplerian velocity
$\vphi=v_{\rm K}(1-\delta)$; and the initial ratio of the
kinetic energy of the SL along $\theta$ coordinate and
it's thermal energy $\Theta \equiv (\mu m_{\rm p} v^2_{\theta_0})/k\TS $.

As was demonstrated by IS99, the solution of   equations~(\ref{eq:cn})--(\ref{eq:en})
depends  very little  on $\theta_0$ (if $\theta_0$ sufficiently
small  $<0.1$, see below) and $\delta$, but strongly depend on
parameter  $\Theta$.  We choose the solutions which are   closest to the critical
solution (in this solution $\vtheta$ is equal to the sound speed at the
maximum latitude of the spreading layer), but slightly subsonic.  The
necessary critical value of  parameter  $\Theta$  is found by the bisection
method.

\begin{figure}
\centerline{ \epsfig{file=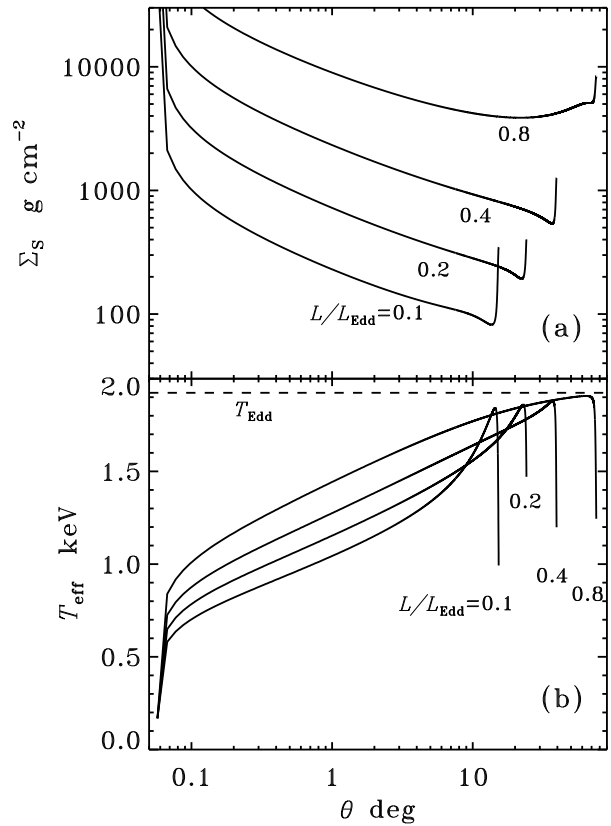,width=8cm}}
\caption{\label{fig3}
The distributions of the surface density $\Sigmas$ and the effective
temperature $\Teff$ over the latitude $\theta$ for the SL
models with different luminosities. Initial SL latitude is $\theta_0 =
10^{-2}$. }
\end{figure}

The distributions of   $\vtheta$, the effective
temperature $\Teff$, and the surface density $\Sigmas$ along the
latitude $\theta$ for four models with the same accretion rate,
corresponding to first model luminosity  are shown in  Fig.~\ref{fig2}.
The first model (shown by the solid
curves) is our model with the pseudo-Newtonian potential and solar hydrogen
abundance $X=0.7$.  The dashed curves are for our model
with GR correction, but with pure hydrogen $X=1$; the dotted curves correspond to
our model without GR corrections
($\Rs=0$ in the equations) with solar hydrogen abundance ($X=0.7$), while the
dot-dashed curves are for the IS99 model with GR corrections and
 solar hydrogen abundance.   It is clear
that the solar abundance lead to a narrower spreading
layer with a smaller surface density.  A higher helium abundance as well as the
the GR corrections lead to a higher effective temperature and a larger 
latitudinal velocity.
Our model gives a slightly wider SL with slightly smaller latitudinal
velocity but same effective temperature and surface density.  Most
calculations below were performed for our model with the GR
correction and solar abundances. The surface density and the
effective temperature distributions along the latitude for models with 0.1,
0.2, 0.4 and 0.8 of the Eddington luminosity are presented in Fig.~\ref{fig3}.
Variations of parameter $\alphab$ lead to some changes in the SL structure. 
The SL column density is inversely proportional to  $\alphab$, 
while the resulting effective temperature decreases only by about 1 per cent with
decreasing of $\alphab$ by an order of magnitude.

\begin{figure}
\centerline{ \epsfig{file=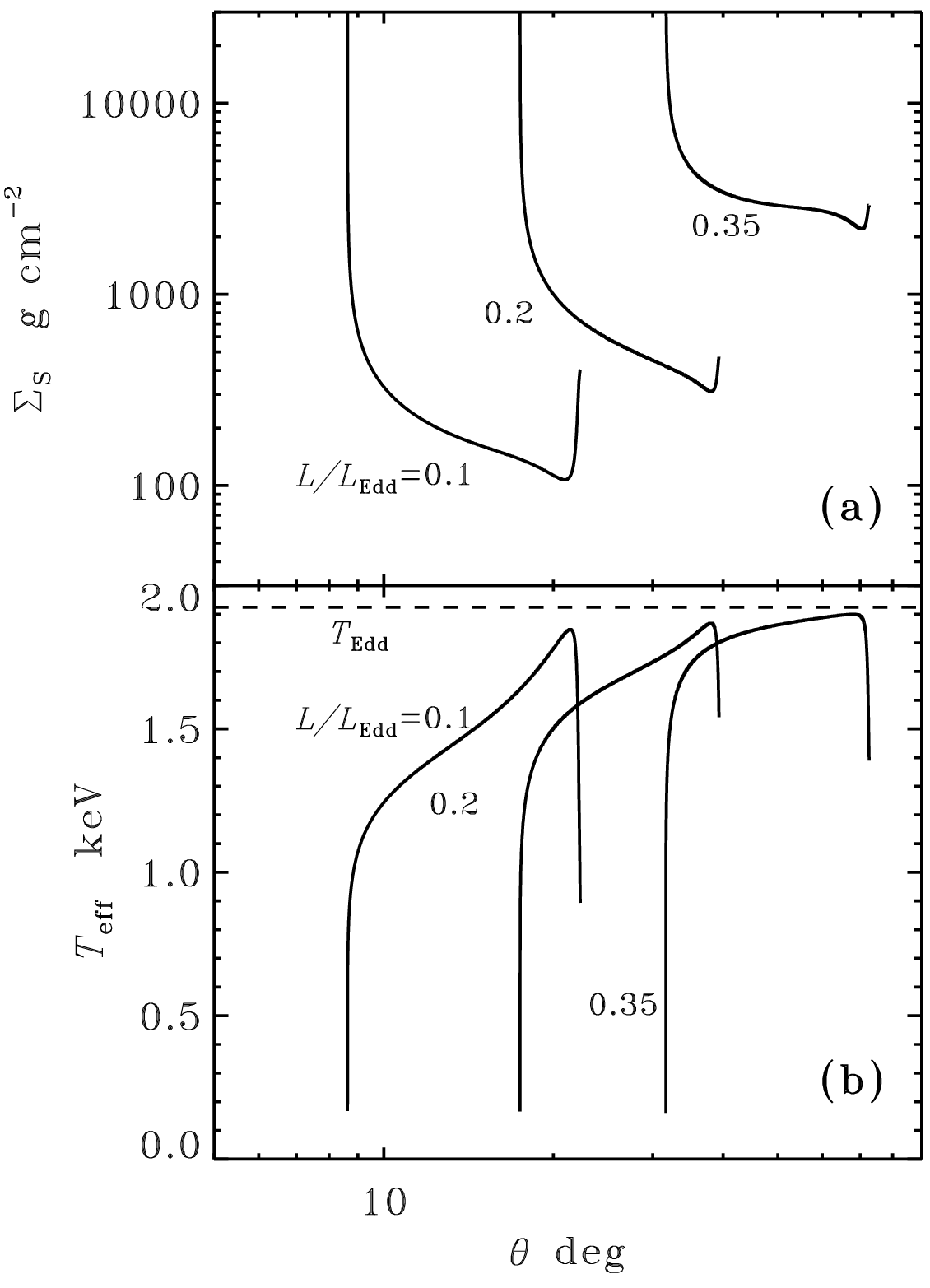,width=8cm}}
\caption{\label{fig4}
The distributions of the surface density $\Sigmas$ and the effective
temperature $\Teff$ along latitude $\theta$ for SL
models with different luminosities. Initial SL latitude $\theta_0 =
\arcsin (0.5 z_0/\Rns)$. }
\end{figure}

The lower boundary of the SL was taken very close
to the equator, $\theta_{\rm 0} \approx 0.01-0.001$, in IS99. Formally,  the
accretion disc thickness is close to zero at the inner boundary, if we take
the usual inner boundary condition for the component of the stress tensor
$W_{r\varphi}(R_{\rm in})=0$. In the case of the accretion disc
around a NS this condition is not correct, and the disc
thickness at the NS surface is considerable. The
luminous accretion disc half-thickness can be evaluated from the
balance of the radiation force and $z$-component of gravity:
\be \label{z0}
      z_0 = \frac{3 \sigmae}{8 \pi c} \dot{M}.
\ee
We calculated the SL models with the initial latitudes
$\theta_{01} = \arcsin (z_0/\Rns)$ and
$\theta_{02} = \arcsin (0.5 z_0/\Rns)$.
The surface density and the effective temperature
distributions along the latitude for the second case are shown in Fig.~\ref{fig4} .
The qualitative behavior of these distributions is close to the case of
small $\theta_0$ with some shift along the latitude. The main
difference is the maximum possible luminosity of the SL. At this luminosity
 the SL reaches the NS poles.
 For $\theta_{02}$, the maximum possible luminosity is about 0.4 $\Ledd$, 
 while for $\theta_{01}$ it is about 0.25 $\Ledd$.

\subsection{Vertical structure of the spreading layer}
\label{sec:vertical}

The IS99 SL model was constructed under an assumption that the
local layer velocity and the radiation flux along height is constant. It
means that the SL is decelerated and energy is liberated in
the infinitely thin  layer at the NS surface. This is an
approximation only and the velocity distribution should not be uniform and
the energy should be generated at all heights. Therefore, we need a more detailed
vertical model of the SL for  calculating its radiation spectrum.

For evaluation of the viscosity parameter $\alphab$,
IS99 used classical theory of
hydrodynamic BLs  \citep{LL59}
with the logarithmic velocity and the energy generation distribution along the height.
In this case,
both the energy generation rate and the velocity gradient are inversely
proportional to the distance from the NS surface $z$
\be
\label{eq:lgr}
\frac{\d q}{\d z} \propto \frac{\d v}{\d z} \propto
   \frac{1}{z}.
\ee
It is clear that these dependencies cannot be correct
in a SL. The SL has a finite thickness with a low
density at the surface. However according to equation~(\ref{eq:lgr}) some
amount of energy has to be generated in the surface layers.

At present time, a theory of the radiation-dominated turbulent boundary
layer does not exist. Thus we here can only make similar assumptions about
the energy generation and velocity gradient along the
height. We assume that these values are inversely proportional to the
surface density measured from the NS surface:
\be \label{eq:qgr}
\frac{\d q}{\d m} = -A~ \frac{q_{\rm 0}}{\Sigmas-m},
\ee
\be \label{eq:vgr}
\frac{\d v}{\d m} = -A~ \frac{v_0}{\Sigmas-m},
\ee
where $A= 2.5 \alphab^{1/2}$, $q_{\rm 0}$ and $v_0$ are the
local radiation flux and the average layer velocity at a given latitude
obtained from the one-dimensional model. Equations (\ref{eq:qgr}) and (\ref{eq:vgr})
 are very close to the IS99 SL
model assumptions (the layer is decelerated and the energy is
generated at the bottom of the layer).
Integration of these equations yields
\be \label{eq:q}
  q(m)=q_{\rm 0} \left[1+A~ \ln\left(1-\frac{m}{\Sigmas}\right)\right],
\ee
\be \label{eq:v}
  v(m)=v_0 \left[ 1+A~ \ln\left(1-\frac{m}{\Sigmas}\right)\right].
\ee
These equations can be used up to some critical column density
\be \label{eq:mstr}
m_*=\Sigmas\left(1-\exp[-A^{-1}]\right),
\ee
which is very close to the local surface density.

The hydrostatic equilibrium equation then
reads
\be \label{eq:hyd}
\frac{\d P_{\rm g}}{\d m} = g_{\rm 0} - \frac{v^2(m)}{\Rns}
-\frac{q(m)}{c}\sigmae
\ee
and the radiation transfer equation is
\be \label{eq:rad}
\frac{1}{3} \frac{\d\varepsilon}{\d m} = \frac{q(m)}{c}\sigmae.
\ee
The temperature and the gas pressure distributions along the
height are:
\begin{eqnarray} \label{eq:tm}
   T(m) &=&   \Teff~ \left[ \frac{3}{4}m \sigmae \
     \left(1-A\left[1+  \frac{\Sigmas-m}{m} \right. \right. \right.
\\ \nonumber
 & \times& \left. \left. \left.  \ln\left(1-\frac{m}{\Sigmas} \right) \right] \right) +\frac{1}{2}\right] ^{1/4}
\end{eqnarray}
\begin{eqnarray} \label{eq:pg}
  P_{\rm g}(m) & = & g_0 m - \frac{v^2_0}{\Rns}  m\left(1-2A+2A^2\right) \\  \nonumber
   & +& \frac{v^2_0}{\Rns}A\left(\Sigmas-m\right) \ln\left(1-\frac{m}{\Sigmas}\right)\\ \nonumber
  &\times& \left[A~\ln\left(1-\frac{m}{\Sigmas}\right)-2A+2\right] \\
\nonumber
&- & \frac{q_0\sigmae}{c}\left[m-A\left(m+\left(\Sigmas-m\right)
\ln\left(1-\frac{m}{\Sigmas}\right)\right)\right].
\end{eqnarray}
These solutions are obtained
 using the boundary conditions at the surface $\varepsilon(m=0)=2q_0/c$ 
 and $P_{\rm g}(m=0)=0$.

At the same time, there is a disagreement between this vertically explicit
model and one-dimensional model, because the velocity and the flux vertical
profiles are different.  We suggest, that the model can be made
self-consistent, if we find a new value of the surface density $\Sigmas'$ 
at a given latitude, which conserves the mass flux
\be \label{eq:msfl}
v_0 \Sigmas= \int_0^{\Sigmas'}  v(m) \d m = v_0
(1-A) \Sigmas'.
\ee
Therefore, the new value of the surface density $\Sigmas' =
\Sigmas/(1-A)$. For $\alphab = 10^{-3}$, this gives
$\Sigmas' = 1.086 \Sigmas$ and we take these values
below for our calculations. There are
similar disagreements for other integrals over the height in equations
(\ref{eq:cont2})--(\ref{eq:energy2}). For example:
\be \label{eq:ke}
v_0^2 \Sigmas= \int_0^{\Sigmas'} v^2(m) \d m =
v_0 (1-2A+A^2) \Sigmas'.
\ee
In this case, we have to take a new value of the surface density
$\Sigmas' = \Sigmas/(1-2A+A^2)$, which gives
$\Sigmas' = 1.171 \Sigmas$ if $\alphab = 10^{-3}$.
Our vertically explicit models disagree
with the one-dimensional ones by about 10 per cent.
Fortunately, the emitted local spectra
depend very little on the surface density of the SL.

\section{Spectrum of the spreading layer}

For calculation of the SL spectra we divide it  into a
number of rings over the  latitude which have different effective
temperatures $\Teff$, matter velocities $v_0$ and surface densities
$\Sigmas$.
We then calculate the vertically explicit model for
each ring, solve the radiative transfer equation and obtain
the local SL spectrum. Then we integrate
local spectra from the SL surface accounting for the general and
special relativity effects.

\subsection{Local spectra}

To calculate a vertically explicit hydrodynamical model with
the radiation transfer  we use standard  methods
for  stellar atmospheres modelling  \citep{M78}. Our models are
obtained in the hydrostatic and the plane-parallel approximations. The
effective temperatures of the considered SL models are rather high
($\sim$ 2 keV) and these models are similar to the atmospheres  of
bursting NSs, where Compton scattering have to be taken
into account.

The vertically explicit local SL model is described by the following
equations: the  equation of
hydrostatic equilibrium (\ref{eq:hyd}), the
energy generation law (\ref{eq:q}), the
velocity law (\ref{eq:v}), the  RTE accounting for the Compton
effect using the   \citet{K57} operator:
\begin{eqnarray}  \label{eq:rtr}
   \frac{\partial^2 ( f_{\nu} J_{\nu})}{\partial \tau_{\nu}^2} =
\frac{k_{\nu}}{k_{\nu}+\sigmae} \left(J_{\nu} - B_{\nu}\right) -
 \frac{\sigmae}{k_{\nu}+\sigmae} \frac{kT}{\me c^2}
\times \\ \nonumber
 x \frac{\partial}{\partial x} \left(x \frac{\partial J_{\nu}}{\partial x} -
3J_{\nu} + \frac{\Teff}{T} x J_{\nu} \left[ 1 + \frac{CJ_{\nu}}{x^3} \right] \right),
\end{eqnarray}
where $x=h \nu /k\Teff$ is dimensionless frequency,
$f_{\nu}(\tau_{\nu}) \approx 1/3$ is the variable Eddington factor, $J_{\nu}$
is  the mean intensity of radiation, $B_{\nu}$ is the black body (Planck)
intensity, $k_{\nu}$ is the opacity due to the free-free and bound-free
transitions, $\sigmae$ is the electron (Thomson) opacity, $T$ is the local
electron temperature, $\Teff$ is the effective temperature of SL at a
given latitude, and $C=c^2 h^2/2(k\Teff)^3$. The optical
depth $\tau_{\nu}$ is defined as
\be
    \d \tau_{\nu} = (k_{\nu}+\sigmae) \d m.
\ee
These equations have to be completed by the energy balance equation
\begin{eqnarray}  \label{eq:econs}
 \int_0^{\infty} k_{\nu}\left(J_{\nu} - B_{\nu}\right) \d\nu -
\frac{1}{4\pi}\frac{\d q}{\d m} -
 \sigmae \frac{kT}{\me c^2} \times \\ \nonumber
\left[ 4 \int_0^{\infty} J_{\nu}
\d\nu - \frac{\Teff}{T} \int_0^{\infty} x J_{\nu}
\left( 1+\frac{CJ_{\nu}}{x^3}\right)  \d\nu \right]=0 
\end{eqnarray}
and by the ideal gas law
\be   \label{eq:gstat}
    P_{\rm g} = N_{\rm tot} kT,
\ee
where $N_{\rm tot}$ is the number density of all particles, as well as
by the particle and charge  conservation laws. We assume local
thermodynamical equilibrium (LTE) in our calculations, so the number
densities of all ionization and excitation states of all elements have been
calculated using Boltzmann and Saha equations.

For solving these equations and computing the local SL model we used
the Kurucz's code {\sc ATLAS} \citep{K70,K93}  modified for high temperature.
All ionization states of the 15 most abundant elements are
taken into consideration. The photoionization cross-sections from the
ground states of all ions are calculated using {\sc phfit2} code \citep{V96}.
For details see \citet{SGS02}   and \citet{I03}. 
The code was also modified to account for Compton scattering.

The scheme of calculation is the following.
First, the input parameters of the
local SL model are defined from the total one-dimensional SL model (see
Sect. \ref{sec:structure}): the effective temperature $\Teff$, the surface gravity $g_0$,
the surface density $\Sigmas$, and the local average layer velocity $v_0$.
Then the analytical vertically explicit model (\ref{eq:tm}--\ref{eq:msfl})
are calculated together with the new value of surface density
$\Sigmas'=\Sigmas/(1-A)$. The calculations are
performed for  the set of 98 column densities
$m$, distributed logarithmically with equal steps from  $m=
 10^{-5}$ g cm$^{-2}$ to $0.99 m_{*}$. The gas
pressure, which is found from equation~(\ref{eq:pg}),
is not varied during the iterations.

For this starting model, all number densities and
the opacities at all depth points and all the frequencies (we use 300
 logarithmically equidistant frequency points) are calculated. The
RTE  (\ref{eq:rtr}) is solved
by the Feautrier method  \citep{M78,ZS91,PSZ91,GS02}
iteratively, because it is non-linear.
Between the iterations we
calculate the variable Eddington factors $f_{\nu}$ and $h_{\nu}$, using
the formal solution of the RTE for three angles.
Usually 5--6 iterations are sufficient to achieve convergence.

We used the usual condition at the outer boundary
\be
    \frac{\partial ( f_{\rm \nu} J_{\nu}) }{\partial \tau_{\nu}} = h_{\nu} J_{\nu},
\ee
where $h_{\nu}$ is the surface variable Eddington factor,
and the inner boundary condition
\be
   \frac{\partial J_{\nu}}{\partial \tau_{\nu}} =
 \frac{\partial B_{\nu}}{\partial \tau_{\nu}}.
\ee
The outer boundary condition is found from the lack of the incoming
radiation at the SL surface, and the inner boundary condition is obtained
from the diffusion approximation $J_{\nu} \approx B_{\nu}$ and $q_{\nu}
\approx 4\pi/3 \times \partial B_{\nu}/\partial \tau_{\nu}$. This condition
is satisfied for any SL optical thickness, because the SL bottom is the
NS surface.
The boundary conditions along the frequency axis are
\be  \label{eq:lbc}
      J_{\nu} = B_{\nu}
\ee
at the lower frequency boundary, $\nu=\nu_{\rm min}=10^{14}$ Hz
($h\nu_{\rm min} \approx$ 0.03 eV $\ll k\Teff$) and
\be  \label{eq:hbc}
x \frac{\partial J_{\nu}}{\partial x} - 3J_{\nu} + \frac{\Teff}{T} x
J_{\nu} \left( 1 + \frac{CJ_{\nu}}{x^3} \right)=0
\ee
at the higher frequency 
boundary $\nu=\nu_{\rm max}=3\ 10^{19}$~Hz ($h \nu_{\rm max}
 \approx$ 100 keV $\gg k\Teff$). Condition (\ref{eq:lbc})
means that at the lowest energies the true opacity dominates the
scattering $k_{\nu} \gg \sigmae$, and therefore $J_{\nu} \approx
B_{\nu}$. Condition (\ref{eq:hbc}) means that there is no photon flux
along the frequency axis at the highest energy.

The solution of the RTE (\ref{eq:rtr}) should also satisfy
the energy balance equation (\ref{eq:econs}) and  the surface
flux condition
\be
    \int_0^{\infty} q_{\nu} (m=0) \d\nu = \sigmasb \Teff^4.
\ee
We calculated the relative flux error along the depth
\be
     \varepsilon_{q}(m) = 1 - \frac{q(m)}{\int_0^{\infty} q_{\nu} (m) \d\nu},
\ee
where $q(m)$ is found from the energy generation law (\ref{eq:q}), and $q_{\nu}
(m)$ is radiation flux at a given depth obtained from the first
moment of the RTE
\be
    4\pi \frac{\partial (f_{\nu} J_{\nu})}{\partial \tau_{\nu}} = q_{\nu}.
\ee
Then the temperature corrections were evaluated using three different
 procedures. The first procedure is the
integral $\Lambda$-iteration method based on the energy
balance equation (\ref{eq:econs}) which was modified to account for
Compton scattering. It works well in the upper layers. The
second one is the modified   Avrett-Krook flux correction, which uses the
relative flux error and is good in  deep layers. And the third one
is the surface correction, which is based on the emergent flux error.
See \citet{K70} for the detailed description of the methods.

\begin{figure}
\centerline{ \epsfig{file=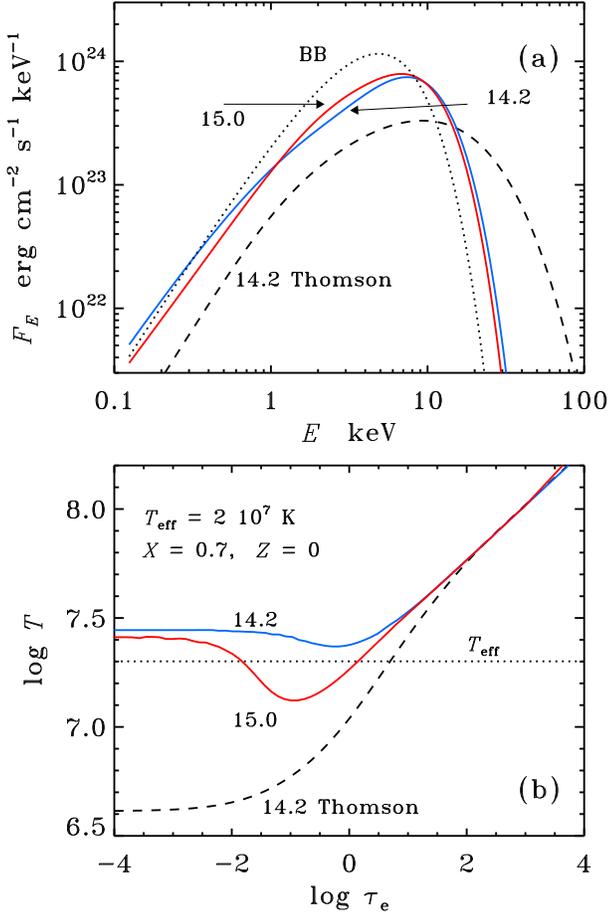,width=8cm}}
\caption{\label{fig5}
(a) The spectra  of  the bursting NS model atmospheres with the effective
temperature $\Teff$ = $2 \ 10^7$ K, solar hydrogen/helium abundances without heavy elements
and $\log~g$ = 14.2 and 15. 
The black body spectrum corresponding to the effective temperature is shown by
the dotted curve.  
(b) The corresponding temperature structure as a function of the 
Thomson optical depth. 
The dashed curves show  the spectrum and the temperature
structure for one model computed not accounting for Compton scattering. 
}
\end{figure}

The iteration procedure is repeated until the relative flux error is
smaller than 1 per cent, and the relative flux derivative error is smaller
than 0.01 per cent. As a result we obtain the self-consistent local SL model
together with the emergent spectrum of radiation.

\begin{figure}
\centerline{\epsfig{file=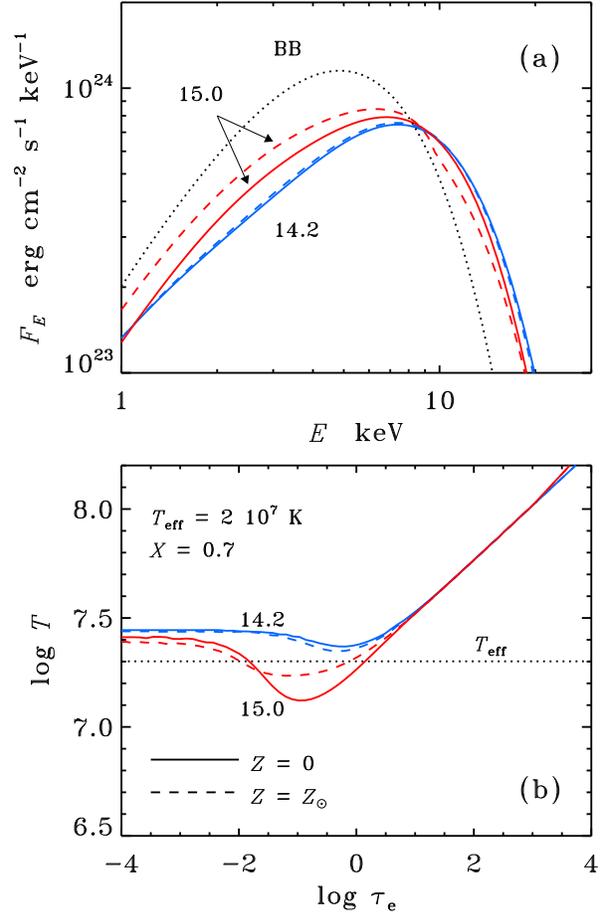,width=8cm}}
\caption{\label{fig6} 
Dependence of the bursting NS model atmospheres 
on metal abundance.
Dashed curves correspond to the solar abundance of metals 
and solid curves to the zero metal abundance. 
Examples are for solar H/He abundances, the same
effective temperature $\Teff$ = $2\ 10^7$ K and
two values of $\log~g$ = 14.2 and 15.
Black body spectrum corresponding to the effective temperature
is shown by dotted curve.}
\end{figure}

Our method of calculation was checked on the atmosphere model of bursting
NS. The equations which describe the bursting
atmosphere are simpler, because there is no velocity field along the
surface ($v_0=0$) and the integral flux is constant along depth
($\d q/\d m=0$).
 We compared our model
atmospheres with the most recent models of \citet*{MJR04}.
The radiation spectra and the temperature structure for
some models with $\Teff=2\ 10^7$~K, solar H/He abundances, and
various surface gravities are shown in Fig.~\ref{fig5}.
These results are in a perfect
agreement with the results of  \citet{MJR04}.  The
emergent spectra and the temperature structure for the models with the
solar abundance of heavy elements are shown in Fig.~\ref{fig6}.

 In the surface layers,  local cooling  is small 
because of the low density, and the temperature equals the Compton temperature of 
radiation which is slightly higher than the effective temperature. 
In dipper layers,  at $\taue\sim 0.1$,
the cooling due to thermal emission (free-free and bound-free) becomes
important (as thermal emissivity per gram  is proportional to density)  
and the temperature decreases. 
At large optical depth the temperature rises again and follows the 
$\taue^{1/4}$ relation,  typical for a grey atmosphere.
At higher surface gravity (at fixed $\Teff$), the plasma density is higher, 
resulting in a more significant temperature dip.
We  see that heavy elements
have rather minimal influence on the models close to the Eddington limit 
(lower~$g$).

\begin{figure}
\centerline{ \epsfig{file=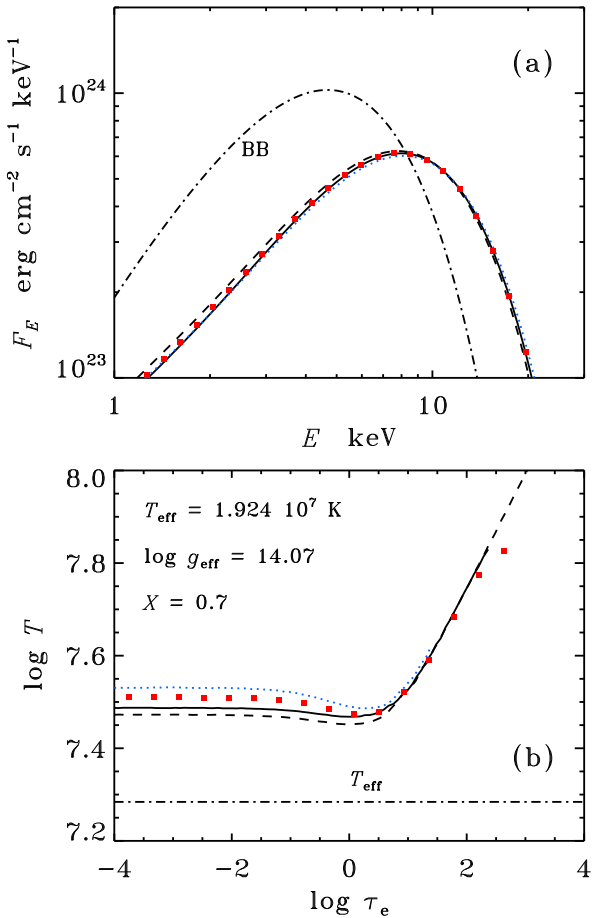,width=8cm}}
\caption{\label{fig7}
Comparison  of the
bursting NS model  (dashed curves) and different local SL models for
the same $\Teff$ and effective gravity $\log~g$.
Solid curves correspond to the local SL model (with the
vertical structure model described in section ~\ref{sec:vertical})
with surface density $\Sigmas= 630\ \mbox{g cm}^{-2}$.
The spectrum and the temperature structure of the local SL model with
the constant velocity and flux derivatives are shown by squares. 
Dotted curves correspond to local SL model (Sect.~\ref{sec:vertical}) 
with $\Sigmas= 63\ \mbox{g cm}^{-2}$.
}
\end{figure}

\begin{figure}
\centerline{ \epsfig{file=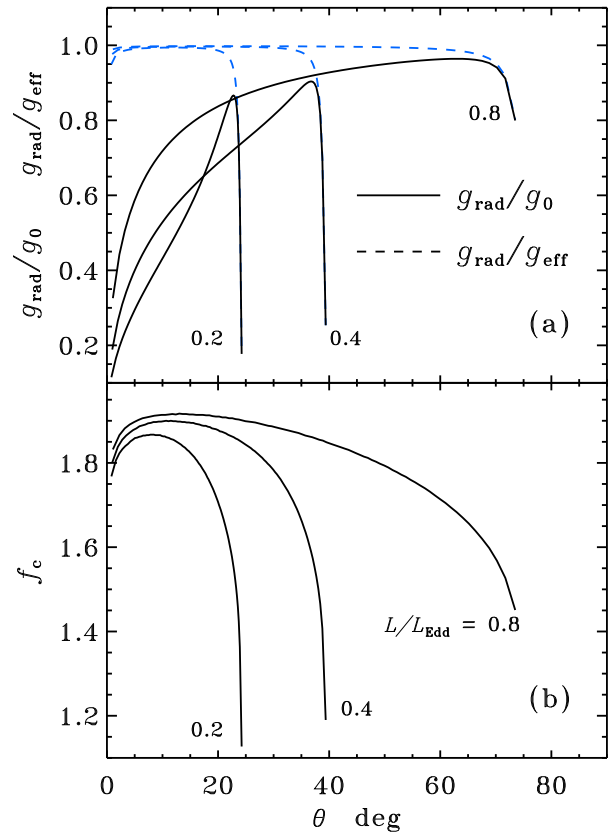,width=8cm}}
\caption{\label{fig8}
(a) Radiative and effective accelerations and (b) the hardness factor $\fc$ (b) as 
functions  of latitude for the SL models of different luminosities $L/\Ledd$. 
}
\end{figure}

The comparison between the bursting NS models and different local SL models for
the same $\Teff$ and effective $\log g$ is shown
in Fig.~\ref{fig7}.
For this SL model we use the vertical structure model, which
is described in Section~\ref{sec:vertical}.  We also investigated, whether the model for the
vertical structure is important for the emergent spectra of local SL models.  We also
calculated the SL model with the constant velocity and flux derivatives:
\be
     \frac{\d v}{\d m} = -\frac{v_0}{\Sigmas}
\ee
and
\be
     \frac{\d q}{\d m} = -\frac{q_0}{\Sigmas}.
\ee
In this case the mass flux conservation requirement
(\ref{eq:msfl}) leads to $\Sigmas'= 2 \Sigmas$.
The spectrum and the temperature structure of this model are shown in
Fig.~\ref{fig7} by squares. The surface temperatures of the local SL
models are higher than the bursting NS model surface temperature. The
reason is the non-zero flux derivative in the energy conservation equation
(\ref{eq:econs}). This means that a part of the energy is released  in the upper
atmosphere and is heating it additionally.  The smaller the surface density
(i.e. the larger the flux derivative), 
the higher the surface temperature. But the
differences  in the temperature structure have very small influence on the
emergent spectra. Therefore we conclude, that details of the vertical structure
have negligible influence on the emergent spectrum for the
optically thick models ($\Sigmas\ge 100\ \mbox{g cm}^{-2}$).

It is well known that the model spectra of bursting NS close to the
Eddington limit
are well described by a diluted Planck spectrum with the color temperature
$\Tc = \fc \Teff$ with the hardness factor $\fc$
varying in the interval 1.6--1.9 and the dilution factor $D=\fc^{-4}$.
 \citet{PSZ91} have derived an analytical formula for the
hardness factor, which successfully describes  high
luminosity ($L \approx \Ledd$) burst spectra:
\be \label{eq:fc}
  \fc = \left( 0.15 \ \ln C_1+0.59 \right)^{-4/5} C_1^{2/15}
\ell^{3/20},
\ee
where $C_1=(3+5X)/(1-\ell)$ and $\ell=L/\Ledd=\grad/\geff$.
Equation (\ref{eq:fc}) works well also for models with heavy elements.

\begin{figure}
\centerline{ \epsfig{file=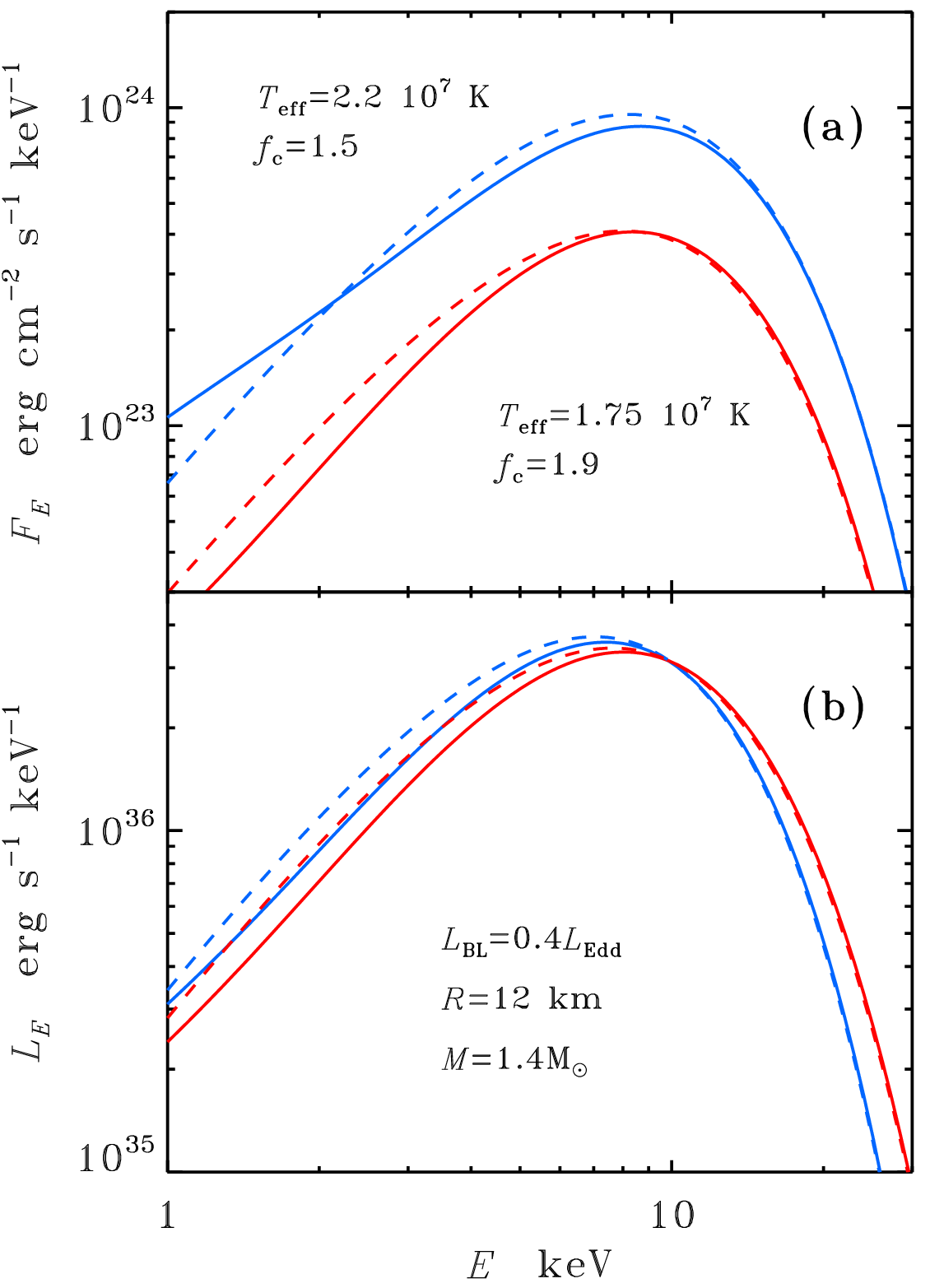,width=8cm}}
\caption{\label{fig9}
(a) Local SL model spectra with high and low effective temperatures (solid curves)
compared to the diluted Planck spectra with $\fc$ given   by equation
(\ref{eq:fc}).
(b) Integral spectra (isotropic luminosities) of the same SL model  (solid curves) 
for different inclination angles $i=0\degr$ (softer spectrum) and $90\degr$ (harder
spectrum) compared to the total spectra obtained by integration of the
diluted Planck spectra  (dashed curves).}
\end{figure}

The local spectra of the optically thick SL (with $L > 0.2 \Ledd$)
are very similar to the burst spectra with corresponding parameters (see
Fig. ~\ref{fig7}). The local SL are very close to the Eddington limit
\be \label{eq:grgeff}
  \grad = \frac{q_0}{c}
  \sigmae \approx g_0 - \frac{v_0^2}{\Rns} = \geff.
\ee
For example, the distributions of the ratios $\grad/g_0$ and
$\grad/\geff$ along the latitude for SL models with three different
luminosities are shown in Fig.~\ref{fig8}a. Corresponding hardness
factor distributions are shown in Fig.~\ref{fig8}b.
The comparison of the two local SL spectra 
(close to equator and at higher latitude) with the diluted
Planck spectra and hardness factors given by equation (\ref{eq:fc})
are shown in Fig.~\ref{fig9}a. 
Closer to the equator, effective gravity is low as centrifugal force 
is large. The gas is levitating above the NS and $\ell$ is close to
unity. The energy dissipation and the effective temperature are low. 
Thus, $\fc$ is large and the spectrum is close to the diluted Planck. 
At higher latitude, the layer is decelerated, while the energy dissipation and $\Teff$ 
grow. However, the effective gravity grows faster reducing $\ell$ and the color correction 
$\fc$. The spectrum shows deviations from the  diluted Planck spectrum at low energies.
At high energies, the Wien part of both spectra can
well be described by the diluted Planck.

\subsection{Integral spectra}

Now we can compute the integral
total model spectrum of the SL, which is seen by a distant observer
accounting for  the relativistic effects such as gravitational
redshift, light bending, relativistic Doppler effect and aberration.
We take into account only half of the SL
because another half is hidden by the accretion disc and 
divide the SL surface on  10 latitude
rings and on 100 angles in azimuth.
In a spherical coordinate system, where the accretion disc coincides with the
$\theta = 0\degr$ plane, the spectrum of the SL is
   \citep{PG03}
\be \label{eq:totsp} 
   F_{\rm E} = \frac{\Rns^2}{D^2} \int\limits_0^{\theta_{\rm SL}}
 \int\limits_0^{2 \pi} \eta^3 \delta^3
I(E',\cos \alpha', \theta) \cos \alpha' 
\cos \theta\ \d\theta\  \d\varphi.
\ee

Here the observed and the emitted photon energies are connected by the
relation $ E=E' \ \eta\ \delta$, where $\eta = \sqrt{1-\Rs/\Rns}$,
the Doppler factor $\delta = 1/\gamma (1-\beta \cos\xi)$,
$ \beta=\vphi(\theta)/c$ (here we neglected low latitudinal velocity), 
the Lorentz factor $\gamma =1/\sqrt{1-\beta^2}$,  and
$ \cos\xi = - \sin\alpha\  \sin i\ \sin\varphi/\sin\psi$. 
The light bending is accounted for by the relation \citep{B02}
\be
\cos\alpha =    \frac{\Rs}{\Rns} + \eta^2 \cos\psi,
\ee
where $\cos\psi = \cos i \sin\theta + \sin i \cos\theta \cos\varphi$,
and the relativistic aberration gives $ \cos\alpha' = \delta\cos\alpha$
\citep{PG03}.
Here $i$ is the inclination angle of the NS polar
axis to the line of sight, $D$ is the distance to the observer, and
$\theta_{\rm SL}$ is the SL boundary. Only visible surface elements
with $\cos\alpha > 0$ give  contribution to the total spectrum.

\begin{figure}
\centerline{ \epsfig{file=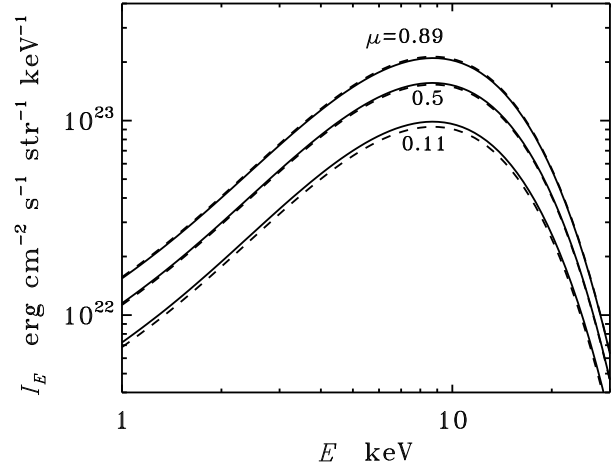,width=8cm}}
\caption{\label{fig10}
Specific intensities 
of the emerging radiation of the local SL model (here $\mu=\cos\alpha'$) 
with the same parameters as in Fig.~\ref{fig7} computed exactly (solid curves)
and using formula (\ref{eq:ang}).
}
\end{figure}

The emitted specific intensity $I(E',\cos\alpha', \theta)$ is taken
from the computed local SL flux assuming angular dependence for the
electron scattering atmosphere
  \be \label{eq:ang}
  I(E',\cos\alpha', \theta) =
  \frac{q_{\rm E'}(\theta)}{\pi} (0.4215+0.86775 \cos\alpha').
  \ee
This formula gives a good approximation to the specific intensity
of the emergent radiation (see Fig.~\ref{fig10}).

\begin{figure}
\centerline{ \epsfig{file=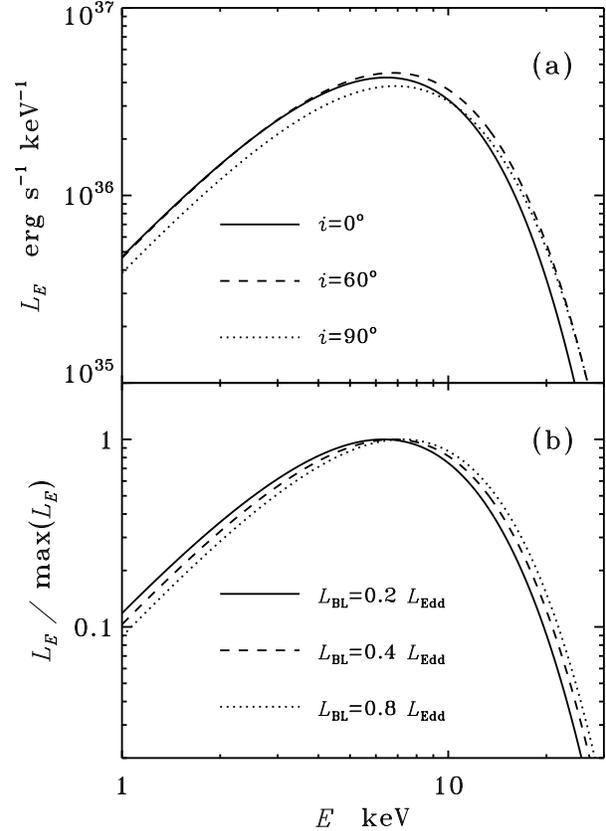,width=8cm}}
\caption{\label{fig11}
Dependence of the SL integral spectra
on the inclination angle $i$ to the line of sight (a)
and on the luminosity (b).}
\end{figure}

The total spectra of the SL model for two inclination
angles, $i=0{\degr}$ and $90{\degr}$, are shown in Fig.~\ref{fig9}b.
The   spectra computed using the local
diluted Planck spectra are shown also for comparison.
The difference is very small in the high energy part ($E> 10$ keV) and more
significant at lower energies ($E < 7$ keV).

Dependence of spectral shape on the inclination angle is not significant (see
also Fig.~\ref{fig11}a). Differences between the SL spectra, which are seen at
different inclinations are comparable to the differences due to change in
the SL luminosities (see Fig.~\ref{fig11}b). It is interesting, that the
total spectra can  also be well described by the diluted Planck spectrum.

The color temperature depends slightly on the assumed turbulence parameter $\alphab$. 
Decreasing  $\alphab$ by an order of magnitude increases $\Tc$ by 0.1 keV.

\section{Comparison with observations}
\label{sec:obs}

In LMXRBs a weakly magnetized NS is surrounded by the accretion disc which
transforms to the boundary/SL close to the NS surface.
At present, about 100 LMXRBs are known.
They can be divided into two different classes.  The Z-sources
are very luminous ($L \sim 0.1 - 1 \Ledd$) and have relatively soft, 
two-component spectra.  Both components are close to the black body with color
temperatures of about 1 keV and 2--2.5 keV.
The atoll sources are less luminous ($L \sim 0.01 - 0.05 \Ledd$) and are
observed in two states, the high/soft and the low/hard.
In the soft state, the radiation spectra  are similar to those of the Z-sources,
while  in the hard state they are close to the spectra of the Galactic
black hole sources in the hard states \citep[e.g. Cyg X-1, see e.g.][]{P98,B00}.
These hard spectra are well
described by unsaturated Comptonization of soft photons in the hot ($kT
\sim$ 30 -- 100 keV) optically thin ($\taue\sim $ 1) plasma.

\begin{figure}
\centerline{ \epsfig{file=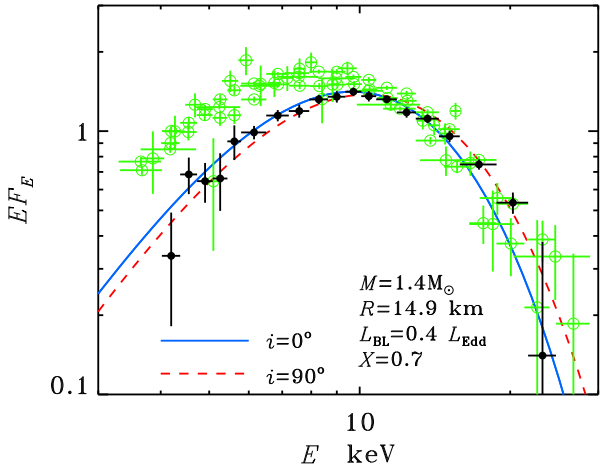,width=8cm}}
\caption{\label{fig12}
Comparison of the observed spectra of the BLs of 
LMXRBs  obtained by the Fourier-frequency resolved spectroscopy  
and the model  spectra of the SL at two
inclination angles to the line of sight. 
Filled circles give the spectrum of GX 340+0 in the normal branch 
\citep[see Fig. 12 in][]{GRM03}, open circles correspond to the spectra 
of five Z- and atoll sources \citep[from][]{RG06}.
}
\end{figure}

The soft component can be associated with the radiation from the accretion
disc, while the hard one with the boundary/SL \citep{M84} or possibly with 
a corona or hot optically thin inner accretion flow \citep[see discussion in][]{DG03}
in case of low-luminosity atoll sources. At high luminosities, the BL 
is   optically thick and its effective temperature is higher  than that of the accretion disc,
because the BL is smaller than the accretion disc, while their luminosities are 
comparable. The hard component is also more variable than the soft component 
at the timescales from millisecond to 1000 seconds  \citep*{M84,GRM03}.
The Fourier-frequency resolved spectroscopy confirms that a component variable 
at high frequencies (and sometimes showing quasi-periodic oscillations, see 
\citealt{vdK00})  has a blackbody-like 
spectrum with the color temperature $\Tc= 2.4\pm 0.1$ keV \citep{GRM03,RG06} 
which is very similar for the five investigated sources.
On the other hand, the variability of the soft component is very similar to the variability of the
soft component of black hole sources in their soft states, which is associated with the accretion disc. 
Based on these arguments, we associate the hard blackbody-like component with the BL
and compare  our theoretical SL spectra with it.

Spectra computed for one SL model together with the observed 
BL spectra obtained by the Fourier-frequency resolved spectroscopy 
 \citep{GRM03,RG06} are shown in Fig.~\ref{fig12}.  
 We see a very good  agreement between theoretical spectra and the 
 spectrum of GX 340+0 at the normal brach (at high accretion rates). 
 The spectra of five Z- and atoll-sources (open circles) are 
 similar to our SL spectra at high energies, but have a soft excess.  
 This excess may be related to the emission of the classical BL, 
 the inner part of the  accretion disc. 
 The observed spectral similarity gives us a confidence to try 
 to determine NS parameters from the observed spectra.

\begin{figure*}
   \begin{center}
     \begin{minipage}[c]{10.5cm}
      \epsfig{file=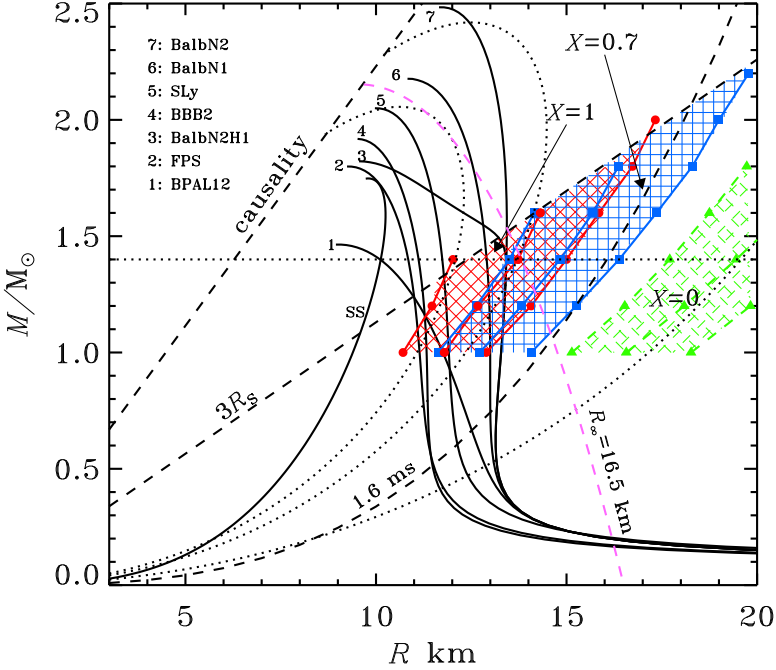,width=\textwidth}
    \end{minipage}\hfill
    \begin{minipage}[c]{6cm}
\caption{\label{fig13}
Permitted region (shaded) for the NS mass and radii, which can have
SLs with color temperatures $2.4 \pm 0.1$ keV \citep{GRM03,RG06} 
at luminosities similar to those observed in the Z-sources.
The left  shaded region (with boundaries  and the 
centre marked by circles) 
is for the accreting matter containing hydrogen only, the
middle one (boundaries and centre marked by squares) is for the solar composition, and the right 
one (marked by triangles) is for pure helium.
The left boundary of each region corresponds to $\Tc=2.5$ keV, 
while the right boundary gives 2.3 keV.
Dotted curves correspond to a simple estimation of $M-R$ relation
given by equation (\ref{eq:eos}) with $\fc=1.7, \ell=0.8$ for the
three chemical compositions. 
Various theoretical mass-radius relations for neutron and strange stars are
shown for comparison  \citep*{HPY06}.
Dotted horizontal line corresponds to the NS mass of $1.4 \msun$. 
The almost vertical dashed curve corresponds to the thermally emitting NS 
with  the apparent   radius of $R_{\infty}=16.5$ km 
(see eq. [\ref{eq:rmrinf}] and \citealt*{T05}).
The lower limit  on the NS mass as a function of radius for a given rotational period 
$P$ (in ms) as derived from the NS stability $M/\msun>
0.865 (R/10\ \mbox{km})^3 P_{\rm ms}^2$ \citep{LP04} 
is presented for $P=1.6$ ms. }
    \end{minipage} \hspace*{0.5cm}
  \end{center}
\end{figure*}

As we have shown above the spectrum of the SL can be 
represented by a diluted blackbody. The effective temperature of radiation
is determined by the critical temperature from equation (\ref{eq:tc}),
where the left-hand side is multiplied by $\ell$, the ratio of the local flux to the 
critical Eddington one (reduced due to the action of the centrifugal force).
The  observed color temperature is $\Tc=\fc \sqrt{1-\Rs/\Rns}\  \Tcr$,
where corrections are made for spectral hardening and gravitational redshift.
For the known color correction and $\ell$, the NS radius as a function of 
compactness $\Mns/\Rns$ can then be found from 
\be \label{eq:eos}
\Rns=       \frac{ \ell \fc^4 c^3} { 2 \sigmasb \Tc^4 \sigmae}   
\frac{\Rs}{\Rns}  \left( 1- \frac{\Rs}{\Rns} \right) ^{3/2} .
\ee
 Assuming $\fc=1.6-1.8$ and $\ell=0.8$, \citet{RG06} obtained 
constraints on the NS mass-radius relation (shown in Fig. \ref{fig13} by dotted curves). 
The maximum NS radius is reached for $\Rs/\Rns=2/5$: 
\be \label{eq:rmax}
\Rns_{\max}=     \frac{24.6}{1+X}      \frac{ \ell}{0.8}   \left( \frac{ \fc}{1.7} \right)^4 
 \left( \frac{ \Tc}{2.4\ \mbox{keV}} \right)^{-4} \ \mbox{km} .
\ee

Here instead we calculate exactly a grid of the SL model spectra, where the main input
parameters are the NS mass $\Mns$ and radius $\Rns$, and
the SL  luminosity.  The NS mass is varied from 1 to 2 $\msun$
with a step of 0.2 $\msun$, and the NS radius is varied from 10 to 24 km with
a step 1 km.  Only the models with $\Rns > 3 \Rs$ are considered.
We take $\alphab=10^{-3}$ and luminosity of $0.4 \Ledd$, and compute 
spectra  for four inclination angles 0, 30, 60 and 90 degrees and for three 
chemical compositions: pure hydrogen ($X=1$), solar
abundance ($X=0.7$), and pure helium ($X=0, Y=1$).
The spectra  are fitted by the black body and the corresponding 
color temperatures are found.

Models with higher He abundance have a smaller hardness factor as can be 
seen from  equation (\ref{eq:fc}). However, the local effective temperature of the SL is 
higher for  larger He abundance (see eq. \ref{eq:tc} and Fig. \ref{fig2}c). 
The higher $\Teff$ leads to a higher color temperature of the integral SL spectrum.
For example, at NS radius of 13 km and mass $1.4\msun$ 
pure hydrogen  models give color temperature of about 2.5 keV, while 
pure helium models produce harder spectra with $\Tc\approx 3$ keV.

Contours corresponding to the color temperature equal 2.3 (right), 2.4
(central) and 2.5 keV (left) are shown on the $\Mns-\Rns$
plane (Fig.~\ref{fig13})  together with the NS models for
various equations of state. These iso-temperature curves are
shown for the inclination angle $i=45{\degr}$.
Comparison of the observed spectra to the theoretical spectra
of the SL constrains the NS radius at $13.5 \pm 1.5$ km (for pure
hydrogen $X=1$ model), $14.8 \pm 1.5$ km (solar composition $X=0.7$)
and $19 \pm 1.5$ km (pure helium $X=0, Y=1$) assuming the NS mass of
1.4 solar mass. For pure hydrogen and solar abundance, the permitted
radii are consistent with the hard equation of state of the NS matter. 
If the composition is solar, but the heavier elements are able to sink,
the emitted spectra would correspond to a pure hydrogen atmosphere 
requiring thus smaller radii. 

Increasing the inclination to $90{\degr}$ increases the deduced NS radii by about 10 per cent, 
while assuming  $i=0{\degr}$, gives a 15 per cent  reduction on $R$. 
The uncertainty in the luminosity increases  the width of possible NS radii by about 50 per cent.
Another source of uncertainty comes from the  turbulence parameter $\alphab$. 
With $\alphab$ decreasing by an order of magnitude the spectrum hardens by  0.1 keV. 
This results in  about 15 per cent decrease of the NS radius that is required 
to produce the observed spectra. Thus  $\alphab\sim 10^{-5}$ is needed 
to reconcile the derived NS radii with the soft equations of state (assuming solar 
composition). Such a small $\alphab$ at the same time yields a very large column 
density of the SL and a rather long life-time of the accreting gas in the layer  
(of the order of 1 s, instead of 10 ms as in the model of IS99). 

Finally, we would like to emphasize that our method of determination of the 
NS radius from the SL spectrum is based on the 
observed color temperature of radiation alone, 
because  the SL radiates locally at almost Eddington flux.
The color temperature can be related to the effective temperature which 
 is a function of the stellar compactness (and chemical composition)  as given 
 by equation~(\ref{eq:tc}). 
This method  is identical to that  used for the radius-expansion X-ray bursts which are 
believed to reach Eddington  luminosity \citep*[see e.g.][]{LvPT93}.
In contrast to the standard methods based on the modeling of the thermal emission 
from the NS surface \citep[see for example][]{vpL87,T05}, 
there is no need to know precisely either the area 
of the emitting region, or the distance to the source. 

As the standard method gives the apparent stellar radius at infinity,  
which is related to the NS parameters through 
\be \label{eq:rmrinf}
R_{\infty} = \Rns \left( 1- \frac{\Rs}{\Rns} \right)^{-1/2}  ,
\ee
the allowed band  of $\Rns$ and $\Mns$ is nearly orthogonal to that obtained
from the color temperature and equation (\ref {eq:eos}) 
(see the almost vertical dashed curve 
in Fig. \ref{fig13}).  Thus for a NS, where both the thermal emission from the surface 
(e.g. during the quiescence) and the BL emission (during the accretion phase) 
are observed, it would be possible to determine $\Rns$ and $\Mns$  independently.
Interestingly our constraints on the NS radius are very similar 
to those obtained by \citet{HR05}  for the thermally emitting 
quiescent NS X7 in the globular cluster 47 Tucanae $\Rns=14.5^{+1.8}_{-1.6}$ km. 
They are also consistent with the lower limit $\Rns>14$ km obtained by \citet{T05} 
for the isolated  NS RX~J1856-3754.

\section{Conclusions}

We have derived the one-dimensional   equations   describing the SL model on a spherical 
NS surface from the usual hydrodynamic equations. The obtained equations are 
similar to those in IS99, except for the  energy conservation law  where we neglected
the surface density of the  gravitational potential energy which is of the second order in $H/R$. 
This difference, however,  leads only to small quantative changes. 
We have also implemented a pseudo-Newtonian potential 
to account for the main general relativity corrections  and considered  
various chemical compositions of the accreting matter. 

We have studied the vertical (radial) structure of the SL  with different
assumptions about the vertical distributions of the radiation flux and azimuthal velocity. 
The temperature structure and the emergent radiation spectra of the SL
are computed accounting for the effect of Compton scattering. We showed that
the local (at a given latitude) emergent spectra depend very little on details
of the SL vertical structure in optically thick cases with
$\Sigmas\gtrsim 100\  \mbox{g\ cm}^{-2}$ ($L \gtrsim 0.1 \Ledd$). 
These spectra can be described by the diluted Planck spectrum and are similar
to the spectra  of X-ray bursts  with the same effective temperature and the 
effective surface gravity.

The integral SL spectra were computed accounting for relativistic effects such as
the gravitational redshift and light bending, the relativistic Doppler
effect and aberration. These spectra slightly
depend on the inclination angle to the line of sight and on the SL luminosity.
The local effective temperature increases with latitude, while the hardness factor
$\fc$  decreases. This leads to only slight variation of the color temperature
on latitude. As a result, the integral spectra can also be well 
described by a single-temperature diluted Planck spectrum. 
 
We compared our theoretical integral SL spectra
with the observed  spectra of the LMXRBs BLs. 
The observed color temperature of 2.4 $\pm$ 0.1 keV \citep{GRM03,RG06}
can be reproduced for hard equations of state of NS material.
Our model constrains radii of NSs in LMXRBs to 13--16 km for a 
1.4 solar mass star. Soft equations of state (smaller NS radii) 
can be reconciled with the observed spectra only for very low viscosity 
$\alphab\sim10^{-5}$.
Calculation of $\alphab$ from the first principles is a challenging 
problem that deserves further attention.

\section*{Acknowledgments}
 
This work was supported by 
the Academy of Finland grants 107943 and 102181,
the Jenny and Antti Wihuri Foundation, 
RFBR grant 05-02-17744, and the
Russian President program for support of the leading
science school (grant Nsh - 784.2006.2).
We are grateful to M. Revnivtsev for providing us with the 
spectral data, and to D. G. Yakovlev and P. Haensel for 
the  theoretical  mass-radius relations for neutron and strange stars.
We thank the  referee for useful comments.


\label{lastpage}

\end{document}